\documentclass[sn-nature,Numbered]{sn-jnl}
\usepackage[markup=underlined]{changes}
\usepackage{todonotes}
\setcommentmarkup{\todo[color={authorcolor!20},size=\scriptsize]{#3: #1}}

\usepackage{graphicx}%
\usepackage{multirow}%
\usepackage{amsmath,amssymb,amsfonts}%
\usepackage{amsthm}%
\usepackage{mathrsfs}%
\usepackage[title]{appendix}%
\usepackage{xcolor}%
\usepackage{textcomp}%
\usepackage{manyfoot}%
\usepackage{booktabs}%
\usepackage{algorithm}%
\usepackage{algorithmicx}%
\usepackage{algpseudocode}%
\usepackage{hyperref}       %
\usepackage[utf8]{inputenc} %
\usepackage[T1]{fontenc}    %
\usepackage{url}            %
\usepackage{booktabs}       %
\usepackage{nicefrac}       %
\usepackage{microtype}      %
\usepackage{xcolor}         %
\usepackage[export]{adjustbox}
\usepackage{wrapfig,amsthm,textcomp,amssymb}
\usepackage{colortbl}
\usepackage{pifont}%
\usepackage[capitalise,noabbrev,nameinlink]{cleveref}
\usepackage{adjustbox}
\usepackage{subfigure}
\usepackage{soul}
\usepackage{cancel}
\usepackage{graphicx}
\usepackage{listings}%
\definecolor{good}{HTML}{ffd2ad}
\definecolor{good}{HTML}{74c476}
\setcitestyle{comma,open={},close={},super}
\usepackage{lineno}

\begin{document}

\title[Article Title]{Breaking the Timescale Barrier: Generative Discovery of Conformational Free-Energy Landscapes and Transition Pathways}

\author*[1]{\fnm{Chenyu} \sur{Tang}}\email{chenyu.tang@univ-lorraine.fr}
\equalcont{These authors contributed equally to this work.}
\author*[1]{\fnm{Mayank Prakash} \sur{Pandey}}\email{mayank.pandey@univ-lorraine.fr}
\equalcont{These authors contributed equally to this work.}
\author[1]{\fnm{Cheng Giuseppe} \sur{Chen}}\email{cheng-giuseppe.chen@univ-lorraine.fr}
\author[2]{\fnm{Alberto} \sur{Meg\'ias}}\email{alberto.megias@upm.es}
\author[1]{\fnm{François} \sur{Dehez}}\email{francois.dehez@univ-lorraine.fr}
\author*[1,3,4]{\fnm{Christophe} \sur{Chipot}}\email{chipot@illinois.edu}

\affil*[1]{Laboratoire International Associ\'e Centre National de la Recherche Scientifique et University of Illinois at Urbana-Champaign, Unit\'e Mixte de Recherche n$^\circ$7019, Universit\'e de Lorraine, B.P. 70239, 54506 Vand\oe uvre-l\`es-Nancy cedex, France}
\affil[2]{Complex Systems Group and Department of Applied Mathematics, Universidad Polit\'ecnica de Madrid, Av. Juan de Herrera 6, E-28040 Madrid, Spain}
\affil[3]{Department of Biochemistry and Molecular Biology,  University of Chicago, Chicago, USA}
\affil[4]{Theoretical and Computational Biophysics Group, Beckman Institute, and Department of Physics, 
University of Illinois at Urbana-Champaign, Urbana, USA}

\abstract{Molecular transitions---such as protein folding, allostery, and membrane transport---are central to biology yet remain notoriously difficult to simulate. Their intrinsic rarity pushes them beyond reach of standard molecular dynamics, while enhanced-sampling methods are costly and often depend on arbitrary variables that bias outcomes. We introduce Gen-COMPAS, a generative committor-guided path sampling framework that reconstructs transition pathways without predefined variables and at a fraction of the cost. Gen-COMPAS couples a generative diffusion model, which produces physically realistic intermediates, with committor-based filtering to pinpoint transition states. Short unbiased simulations from these intermediates rapidly yield full transition-path ensembles that converge within nanoseconds, where conventional methods require orders of magnitude more sampling. Applied to systems from a miniprotein to a ribose-binding protein to a mitochondrial carrier, Gen-COMPAS retrieves committors, transition states, and free-energy landscapes efficiently, uniting machine learning and molecular dynamics for broad mechanistic and practical insight.}


\maketitle


Rare transitions between long-lived metastable states are a hallmark of complex dynamical systems ranging from chemical systems, biomolecules, to materials~\cite{yang2009, Lindorff-Larsen2011fold,ajaz_2011,moradi2013,BK01,sipka_differentiable_2023_inproceedings, jung2023machine, kang2024Parrinello}.
In molecular systems, these events underlie conformational changes, chemical reactions, and recognition and association phenomena, central to biological functions. A precise map of the transition-state ensemble (TSE), dominant pathways, and free-energy landscape (FEL) is key to mechanistic insight and advances in drug discovery, protein design, and catalysis.

Traditional strategies for accessing such information often rely on using either brute-force molecular dynamics (MD) \cite{Dror2012} or enhanced-sampling strategies \cite{Chipot2023}, such as umbrella sampling, metadynamics, adaptive biasing force, and related  approaches---which rely on a priori chosen collective variables (CVs) assumed to represent the slow dynamics~\cite{rogal_reaction_2021,branduardi_b_2007,chen_enhancing_2022}.
Brute-force MD can, in principle, provide an unbiased view of the dynamics, but requires enormous computational resources. Specialized supercomputers, such as Anton~\cite{shaw2008anton}, have been developed explicitly to perform such simulations,  yet the scope of these machines remains limited to systems and timescales appreciably shorter than many biologically or chemically relevant processes \cite{Shaw2021}.
Enhanced-sampling schemes attempt to overcome these limitations by biasing the dynamics along a coarse-grained reaction coordinate (RC) described by a small set of CVs. However, their effectiveness hinges on the quality of the chosen CVs. If key coordinates are excluded from the CVs, biasing may misrepresent the mechanism at play, or fail to accelerate sampling. The advent of artificial neural networks (NNs) has enabled novel data-driven approaches for identifying suitable CVs to enhance sampling, e.g., autoencoder-based NNs~\cite{WN18,BGLS21,FAB21} and graph NNs~\cite{Pengmei2025,ZWT25}. 

Still, even when the chosen CVs capture the relevant dynamics, their practical implementation can pose challenges. In particular, an inadequate bias may drive the system away from thermodynamic equilibrium \cite{Miao2021} and potentially distort the molecular structure in unintended ways. To overcome these challenges, efforts to define optimal pathways have led to the development of numerical approaches like the finite-temperature string method~\cite{e_finite_2005,maragliano2006string,fu2020finding} and the string method with swarm of trajectories~\cite{pan_finding_2008}. The resulting zero-drift pathways (ZDPs), possibly minimum free-energy pathways (MFEPs) describe the  transitions in terms of curves embedded in the chosen CV space, along which enhanced sampling can be performed, using path CVs (PCVs) \cite{branduardi_b_2007}. The computational cost associated to path-finding methods is, however, usually prohibitive. 

More recent developments in transition-path theory\cite{VT05, SPNCB19, roux2022transition, jung2023machine, strahan2023predicting} highlight the committor, $q$, as the optimal RC, with committor-consistent strings (CCS)\cite{he_committor-consistent_2022} tracing dynamically meaningful pathways, usually distinct from MFEPs or ZDPs. Variational formulations~\cite{chen_discovering_2023, MCCCRC25, giuseppe_grad_2025} enable learning $q$ from biased PCV sampling using variational committor neural networks (VCN) iteratively, optimizing pathways and sampling the conformational space simultaneously, boosting the the search for transition pathways in the CV space. The committor can also be expressed in the full Cartesian space with NNs, mapping the transition dynamics without predefined CVs~\cite{sergio_qGNN_2025}, with the potential of being a general descriptor of rare events with minimalist a priori knowledge. However, thus far, learning the committor and the CCSs still relies heavily on enhanced-sampling methods. 

In recent years, generative models such as Boltzmann generators~\cite{noe2019boltzmann} and related architectures have been proposed to directly generate equilibrium conformations.
Frameworks like MDGen\cite{jing2024generative} and BioEmu~\cite{lewis2025scalable} claim to directly emulate molecular structures. Although impressive in terms of throughput, these approaches require massive datasets for pretraining or fine-tuning---often hundreds of milliseconds of MD or large-scale experimental measurements---before they can be generalized to qualitatively new systems under specific conditions. For rare events, such as large conformational changes or recognition and association phenomena, collecting sufficient data is extremely difficult. Moreover, because the generated conformations arise from sampling in a learned latent space rather than from the true molecular Hamiltonian, the resulting FELs are not exactly comparable to those obtained from direct MD. Artifacts in the inherent uncertainty of NNs, mode imbalance, and bias toward training-set geometries can easily distort thermodynamic predictions, limiting their reliability for quantitative analysis. Developing a framework that is both physically rigorous and computationally efficient, while faithfully capturing the kinetics and thermodynamics of rare molecular events, remains a daunting challenge.

\section*{Gen-COMPAS framework}

\begin{figure*} 
	\centering
	\includegraphics[width=\textwidth]{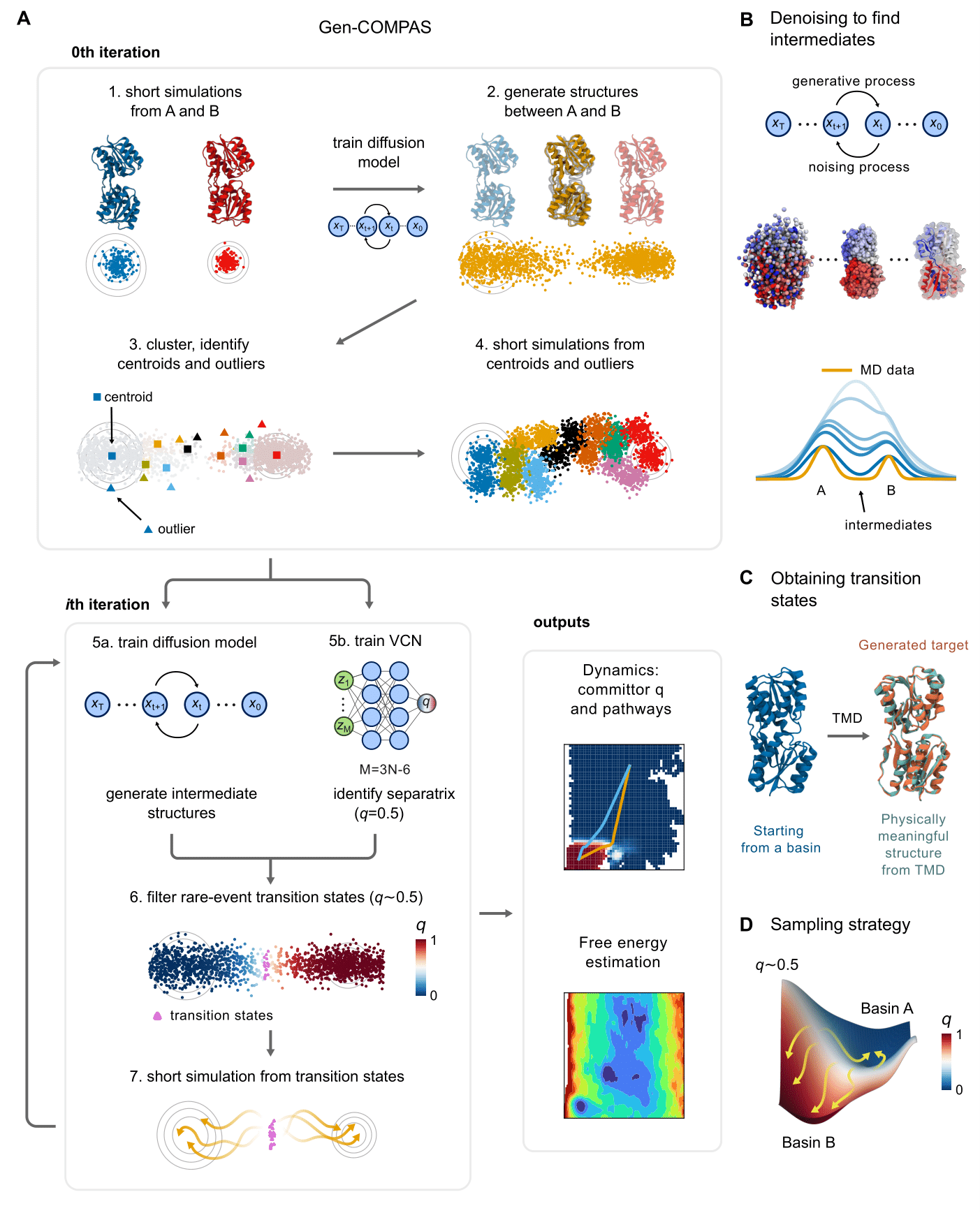} 
	\caption{\textbf{Gen-COMPAS in a nutshell} \textbf{(A)} Overall framework of Gen-COMPAS. \textbf{(B)} Denoising diffusion model: Training and inferring to find the intermediates. \textbf{(C)} Obtaining physically meaningful structures corresponding to transition states using targeted MD (TMD). \textbf{(D)} Overall sampling strategy between two metastable states of Gen-COMPAS guided by the committor ($q$) in a schematic free-energy landscape.}
	\label{fig:framework} 
\end{figure*}

Here, we introduce a generative committor-guided path sampling (Gen-COMPAS) strategy, an iterative framework for exploring rare-event dynamics, which addresses the limitations of a generative model by leveraging the accumulated MD sampling at very reduced computational cost (Fig.~\ref{fig:framework}A). The workflow begins with very short (1--2 ns) unbiased simulations of a few metastable states---typically two, namely $A$ and $B$, identifying the reactant and product states, respectively---to generate an initial dataset for the purpose of training a diffusion-based generative model~\cite{ho2020denoising,song2020denoising}. In turn, this model produces intermediate conformations connecting the states (Fig.~\ref{fig:framework}B), while a high-dimensional committor function, $q$, learned directly in conformational space, identifies near-transition structures with committor values around the separatrix, i.e., the $q=1/2$ hypersurface, which corresponds to an equal probability of returning to either metastable state (Fig. \ref{fig:framework}D).

The intermediates generated by the model provide strategic targets for efficiently navigating the vast conformational space. We then employ two sets of targeted MD (TMD) simulations \cite{Schlitter1994}, initiated from states $A$ and $B$, to converge upon the physically plausible region associated to each target (Fig. \ref{fig:framework}C). From these points lying on the separatrix, we shoot unbiased MD simulations. The newly generated data are subsequently used to train both the diffusion model and the committor predictor for the next iteration, thereby forming a powerful feedback loop that samples the transitional conformations between the two starting metastable states, $A$ and $B$.

Once convergence is achieved, the cumulative sampling enables downstream tasks, including (1) identification of transition states, (2) construction of committor maps projected onto any interpretable CVs defined post-simulation, (3) extraction of CCSs, or transition pathways between $A$ and $B$, and (4) approximation of FELs (see Supplementary Information for further details).

This framework offers several key advantages over existing enhanced-sampling and generative-modeling strategies. By obviating the need for predefined CVs, it eliminates possible biases in FELs or kinetic pathways, which is a common artifact in CV-dependent methods. Sampling is focused on the region of the separatrix, enabling accelerated exploration of transition states compared to brute-force MD. The framework is natively GPU-optimized, ensuring scalability to large biomolecular systems. Critically, while many prior generative approaches~\cite{jing2024generative, lewis2025scalable, costa2025accelerating} are restricted to protein-only systems that exclude ligands or any molecular substrate, Gen-COMPAS is able to explicitly model protein-ligand complexes and other heterogeneous biomolecular assemblies without need for adaptation. Furthermore, our framework enables a direct, simultaneous recovery of kinetic and thermodynamic observables from physically grounded simulations, in contrast to generative models that approximate such properties through latent-space reconstruction. The sampling provided by Gen-COPMAS is therefore physical and generalizable.

The performance of the Gen-COMPAS framework is demonstrated across biomolecular processes of increasing complexity, from $N$--acetyl--$N'$--methylalaninamide (NANMA) and trialanine isomerization in vacuum, to the fast-folding reversible folding of Trp-cage and binding-upon-folding of the ribose-binding protein (RBP) in explicit solvent, to finally the conformational transition of the mitochondrial ADP/ATP carrier (AAC) embedded in a membrane environment. The conformational equilibrium of NANMA and trialanine serve as proof-of-concept systems, for which Gen-COMPAS has provided accurate estimation of FEL, committor map, TSE, and transition pathways in remarkable agreement with earlier work~\cite{chen_chasing_2023, chen_companion_2022,MCCCRC25,sergio_qGNN_2025,giuseppe_grad_2025}, and can be found in the Supplementary Information. 

\section*{Results}\label{sec2}

\subsection*{Fast-folding proteins: Trp-Cage}\label{sec3}

To benchmark the performance of Gen-COMPAS, we have applied this framework to the well-characterized Trp-cage miniprotein. Fast-folding proteins like Trp-cage are central to folding studies, as their simple topology and short folding times offer clean benchmarks for validating computational methods~\cite{Zhou2003_FEL_TrpCage,Marinelli2009_KineticModel_TrpCage,Sidky2019_SRV_MSM_TrpCage}. They are small enough to be simulated with atomistic detail, yet complex enough to capture essential aspects of protein-folding processes such as cooperative transitions and the formation of native-like intermediates. By focusing on such systems, one gains mechanistic insights into folding principles that extend to larger proteins, while stress-testing whether a computational method can reproduce both thermodynamic and kinetic observables. 

\begin{figure*}
	\centering
	\includegraphics[width=\textwidth]{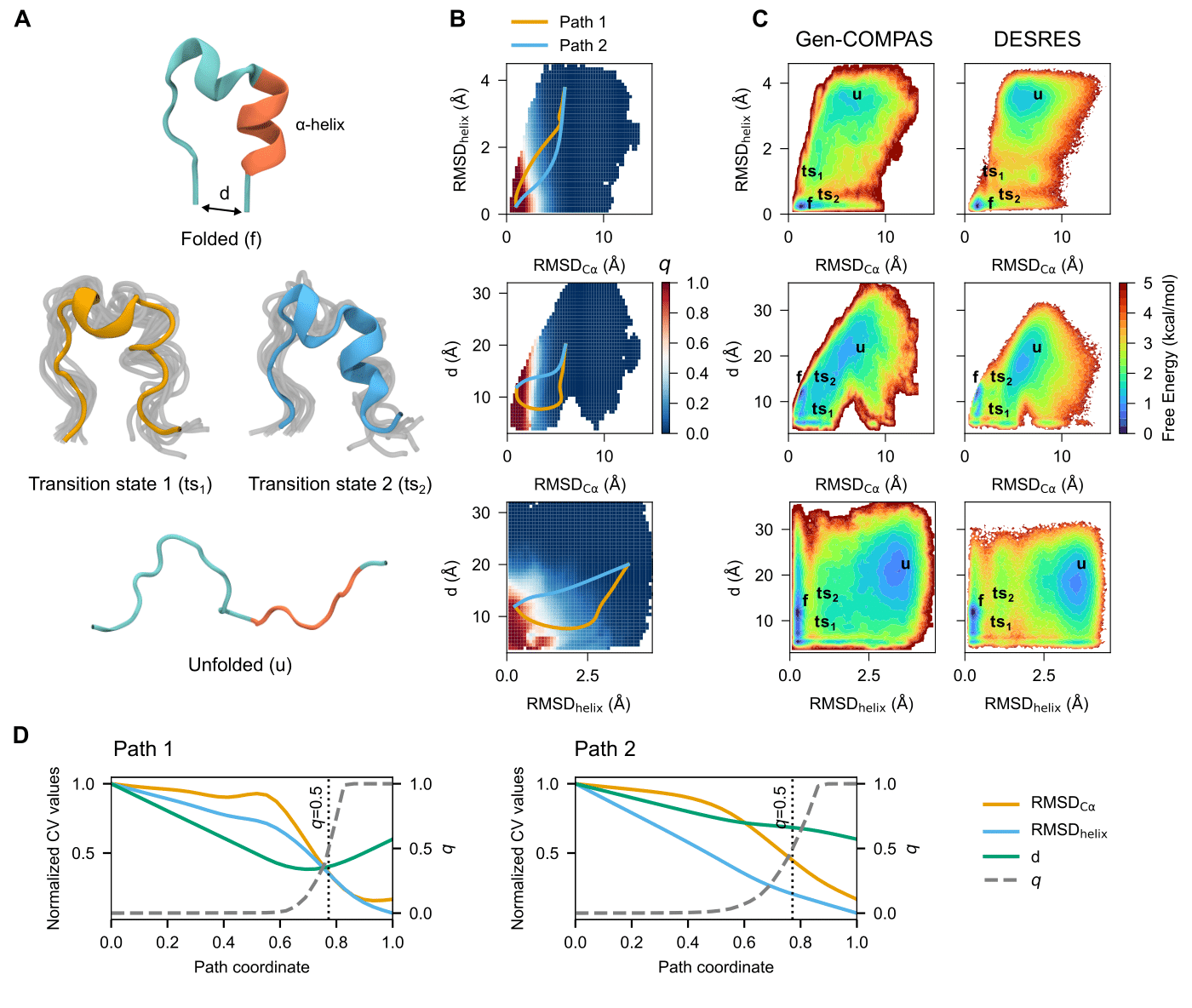} 
	\caption{\textbf{Gen-COMPAS of Trp-cage fast-folding protein} \textbf{(A)} Representative structures of Trp-cage in the folded, transition and unfolded states. The $\alpha$-helix and the end-to-end distance, d, are highlighted. \textbf{(B)} Learned committor ($q$) projected onto the distance root mean square deviation (RMSD) with respect to all the  C$_{\alpha}$ atoms, the RMSD with respect to the C$_{\alpha}$ atoms of the $\alpha$-helix, and the end-to-end distance. The two sampled pathways are also depicted. \textbf{(C)} Free-energy landscape projected onto the same collective variables (CVs) obtained by Gen-COMPAS and by using the DESRES simulations \cite{Lindorff-Larsen2011fold} \textbf{(D)} Normalized values of the CVs and corresponding committor values along the pathways.
    }
	\label{fig:trpcage} 
\end{figure*}

Starting from the folded and unfolded states, Gen-COMPAS successfully identifies the intermediate transition states (Fig. \ref{fig:trpcage}A) and recovers the folding FEL, as illustrated in Fig. \ref{fig:trpcage}C. The estimated free-energy difference is in quantitative agreement with reference data from the $\mu$s-timescale DESRES simulations~\cite{Lindorff-Larsen2011fold}. Analysis of the committor and TSE (Fig. \ref{fig:trpcage}B) further reveals that the folding mechanism is bifurcated, proceeding along two dominant, competing pathways (Fig. \ref{fig:trpcage}D). One route involves early helix nucleation followed by core consolidation. The other route initiates with the hydrophobic collapse of key tertiary contacts around the central tryptophan residue~\cite{Zhou2003_FEL_TrpCage}, with helix formation lagging behind. This observation of multiple folding routes is consistent with previous transition-path sampling studies in explicit solvent~\cite{Juraszek2006}.

It is also noteworthy that Gen-COMPAS reduces the required sampling time from 208 $\mu$s to 594 ns, making it approximately hundreds times more efficient than conventional simulations. Crucially, this efficiency extends beyond equilibrium FELs to kinetic investigations. By enabling direct learning of the committor, the probability of folding before unfolding, Gen-COMPAS characterizes folding kinetics without the need for prohibitively long simulations.

\subsection*{Binding upon folding of the ribose-binding protein}

The RBP is a periplasmic binding protein (PBP), essential for bacterial nutrient uptake, mediating high-affinity ribose recognition and delivery. Like other PBPs, the RBP undergoes a large conformational change between open (apo, or ligand-free) and closed (holo, or ligand-bound) forms~\cite{ravindranathan_ribose_2005,Li2013_RBP_ENM}. Ribose binding is tightly coupled to folding of flexible regions, making the RBP a model for binding-upon-folding processes that illustrate how local disorder drives recognition~\cite{Ren2021_RBP_G134R_Coupling}. Capturing this process is computationally challenging, as it requires simultaneous sampling of large-scale motions, local loop ordering, and ligand docking across a rugged FEL and long timescales.

\begin{figure*} 
	\centering
	\includegraphics[width=\textwidth]{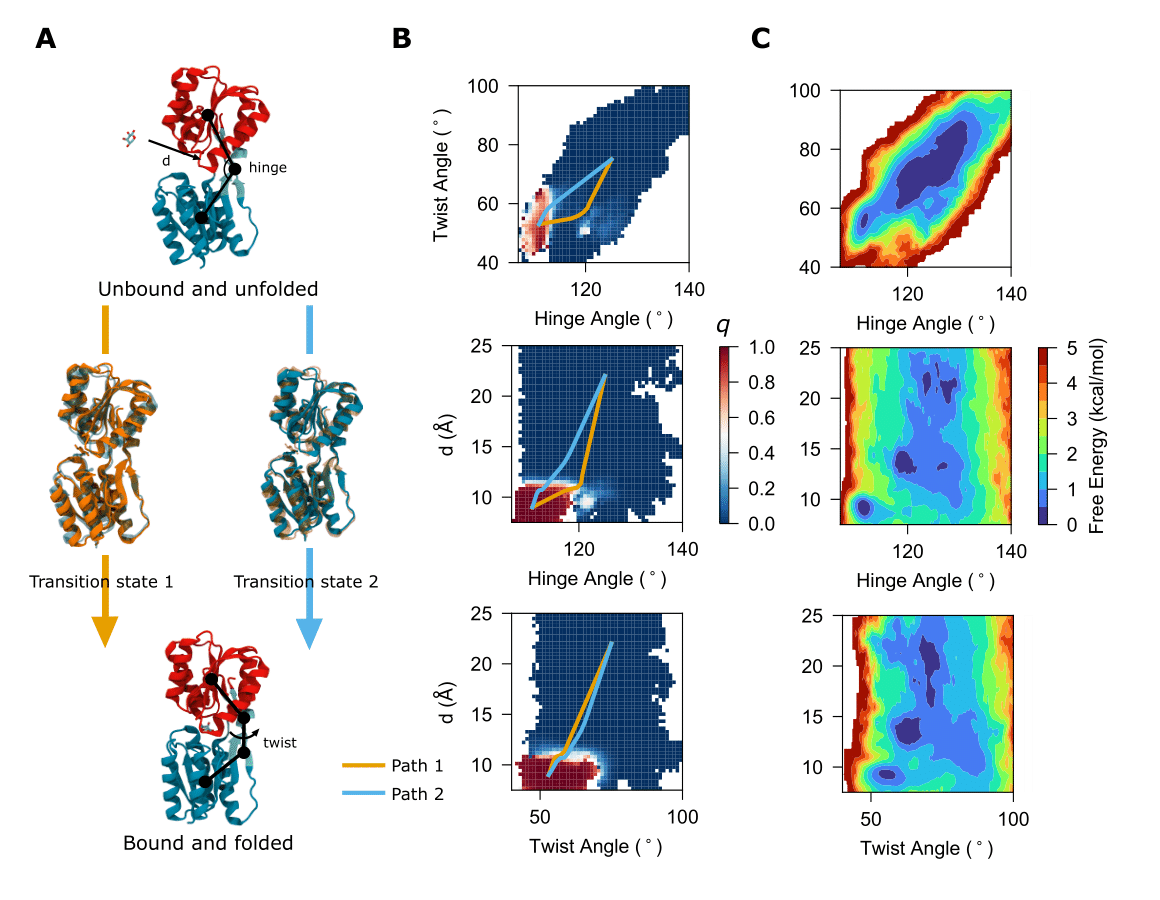} 
	\caption{\textbf{Gen-COMPAS of RBP binding-upon-folding process} \textbf{(A)} Representative structures of RBP-ribose unbound state and bound state, the transition states, and pathways for the RBP binding-upon-folding process. The three collective variables describing the process: d, hinge, and twist~\cite{ravindranathan_ribose_2005}. \textbf{(B)} The committor and committor-consistent pathways revealing the two distinct binding-folding mechanism by Gen-COMPAS. \textbf{(C)} Free-energy landscapes obtained by Gen-COMPAS showing the binding-upon-folding mechanism.
	}
	\label{fig:rbp} 
\end{figure*}

Gen-COMPAS overcomes these challenges, reconstructing the full binding-upon-folding pathway of the RBP. It captures transitions from the disordered open state, through partially folded intermediates, to the final ribose-bound, closed state (Fig.~\ref{fig:rbp}A), revealing the cooperative interplay between ligand binding and protein folding. The binding of ribose to the RBP follows two cooperative pathways (Fig.~\ref{fig:rbp}B). In terms of twist-angle rearrangements, both pathways proceed similarly, indicating a conserved motion. On the other hand, for hinge-bending rearrangements, the pathways diverge: one follows a stepwise induced-fit mechanism, whereby ligand binding precedes protein closure~\cite{Li2021_RBP_LigandInduced_MD,Ren2021_RBP_G134R_Coupling}, while the other exhibits simultaneous binding and folding.

Quantitatively, Gen-COMPAS directly estimates the committor function to identify the TSE, where ribose is interacting with the RBP and the protein is not yet fully closed (Fig.~\ref{fig:rbp}B). Projected FELs along retrospectively defined CVs for ligand position and interdomain angles reveal distinct open and closed basins, separated by barriers consistent with the expected gating motions~\cite{ravindranathan_ribose_2005} (Fig.~\ref{fig:rbp}C). Put together, these analyses unite thermodynamics and kinetics into a coherent mechanistic map of the binding-upon-folding process. By resolving the transition pathways atomistically and quantifying their energetics, Gen-COMPAS offers a general strategy to dissect complex coupled folding-binding events, with broad relevance to other PBPs or intrinsically disordered proteins. 

\subsection*{Unraveling the complex machinery of the mitochondrial ADP/ATP carrier} 

\begin{figure*}
	\centering
	\includegraphics[width=\textwidth]{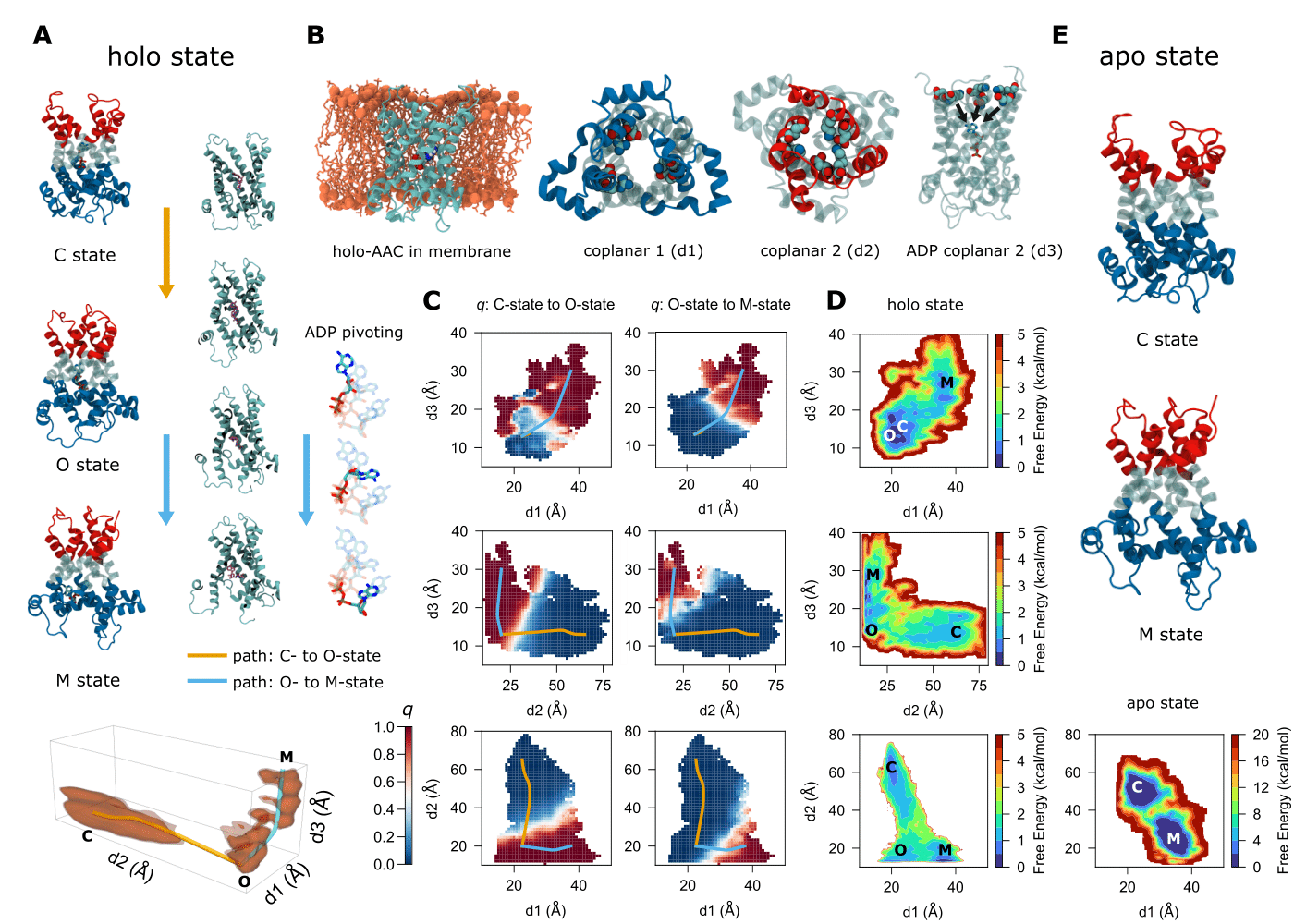} 
	\caption{\textbf{Gen-COMPAS of the mitochondrial ATP/ADP carrier (AAC)} \textbf{(A)} The three meta-stable states, transition states and transition pathway found by Gen-COMPAS from C-state to M-state through O-state of holo-AAC (ADP$^{3-}$-bound). The pivoting motion of ADP$^{3-}$ from O-state to M-state. The conformational transition pathway C $\xrightarrow{}$ O $\xrightarrow{}$ M as well as the basins of the free-energy landscape (FEL) in the three-dimensional CV space. \textbf{(B)} Holo-state of AAC in membrane and the collective variables describing the AAC conformational transition: d1, d2, and d3. \textbf{(C)} Committor and committor-consistant pathway connecting three states and \textbf{(D)} the FEL projected onto two dimensions. \textbf{(E)} The two metastable states of apo-AAC and its FEL projected onto d1 and d2.}
	\label{fig:aac} 
\end{figure*}

The AAC is an essential mitochondrial transporter that maintains cellular energy balance by exchanging cytosolic ADP$^{3-}$ for matrix ATP$^{4-}$ through a strict one-to-one antiport mechanism, fueling processes such as biosynthesis, signaling, and muscle contraction~\cite{klingenberg_aac_2008, mavridou_membrane_2024, kunji_aac_2020}. Because the inner mitochondrial membrane is impermeable to nucleotides, the AAC serves as the exclusive entry point for ADP$^{3-}$, making its function indispensable for oxidative phosphorylation~\cite{senoo_functional_2024}. AAC operates through an alternating-access mechanism, cycling between a cytoplasmic-open (C-) state, which binds ADP$^{3-}$, and a matrix-open (M-) state, which releases ADP$^{3-}$ and binds ATP for export~\cite{tamura_atomistic_2017, skulj_molecular_2020, quadrotta_modeling_2025} (Fig.~\ref{fig:aac}A). While crystallographic studies have captured these end states in an aborted form, employing potent inhibitors, they do not explain how AAC prevents uncontrolled nucleotide leakage during transitions~\cite{springett_modelling_2017}. Biochemical and computational studies have, therefore, proposed a transient occluded (O-) state, in which ADP$^{3-}$ is fully encased within the central cavity, shielded from both sides of the membrane~\cite{pietropaolo_switching_2016, yao_computational_2022}. This state acts as a safeguard to ensure tightly coupled exchange, but direct evidence has remained elusive, as conventional MD struggles to capture this short-lived intermediate, and experimental data have been largely indirect~\cite{yi_molecular_2019, kurauskas_dynamics_2018, senoo_functional_2024}.

With Gen-COMPAS, we are able to overcome these challenges and explicitly demonstrate the presence of the occluded intermediate during ADP$^{3-}$ transport. Our simulations show that the transition proceeds through a well-defined C $\xrightarrow{}$ O $\xrightarrow{}$ M pathway, in which ADP$^{3-}$ first binds steadfastly in the cytoplasmic-open state, becomes trapped in the occluded intermediate through rearrangements of the transmembrane helices, and is finally released into the matrix when AAC adopts the matrix-open conformation (Fig. \ref{fig:aac}A,C). This pathway provides confirmation that the O-state is an obligatory step of ADP$^{3-}$ import to the matrix, rather than an incidental conformation.

The FELs are projected onto three post-simulation interpretable CVs, which are defined as follows. The CV d1 is the sum of distances between the terminal side-chain carbon atom pairs CE–ASP231/CD–LYS32, CZ–ARG137/CE–GLU29, and CZ–ARG234/CD–ASP134; d2 is the sum of distances between the terminal side-chain carbon atom pairs CE–LYS95/CD–ASP195, CE–LYS198/CD–ASP291, and CE–LYS294/CD–ASP92; and d3 is the distance between the N6 atom of ADP$^{3-}$ and the center of mass of the terminal side-chain carbon atoms forming d1 (CE of LYS32, CD of ASP231, CZ of ARG137, CE of GLU29, CZ of ARG234, and CD of ASP134) (Fig. \ref{fig:aac}B). These three CVs capture the key thermodynamics of this transition. Distinct minima corresponding to the C-, O-, and M-states are observed, with barriers consistent with the expected helix-gating motions of AAC~\cite{springett_modelling_2017, pietropaolo_switching_2016} (Fig. \ref{fig:aac}D). Notably, the O-state emerges as a free-energy basin, demonstrating that it is thermodynamically stable enough to act as a genuine intermediate (Fig. \ref{fig:aac}D). Committor analysis further establishes its kinetic role: once the AAC transitions to the occluded state, the probability of progressing toward the matrix-open conformation dominates, confirming that this intermediate is the decisive checkpoint in the inward transport process (Fig. \ref{fig:aac}C).

To further probe the mechanistic differences between ligand-bound and ligand-free conditions, we have conducted Gen-COMPAS for the AAC apo-state (Fig. \ref{fig:aac}D). In stark contrast with the holo-state, where the free-energy barriers are relatively low ($\sim 2.5$ kcal/mol between C- to O-state and $\sim 2$ kcal/mol between O- to M-state), Gen-COMAPS on AAC-APO reveals the disappearance of the O-state and the emergence of a much higher free-energy barrier, reaching $\sim10$ kcal/mol between the C- and M-state. This result confirms that the apo-state transition is strongly disfavored thermodynamically, a finding in agreement with previous studies\cite{pietropaolo_switching_2016}, and highlights the essential role of the substrate in stabilizing the transport-competent pathway~\cite{klingenberg_aac_2008}. 

Beyond elucidating the mechanism that underlies AAC's functionality, findings can link transporter dynamics to mitochondrial physiology and disease. Since AAC is central to bioenergetics, dysfunction in ADP transport can compromise ATP production and contribute to diseases ranging from mitochondrial myopathies to neurodegeneration. By confirming and quantifying the occluded state, Gen-COMPAS provides a mechanistic foundation to understand how mutations or inhibitors disrupt ADP$^{3-}$ transport, paving the way for rational modulation of mitochondrial function in health and pathology.

\section*{Discussion}

Gen-COMPAS is a versatile and efficient framework for exploring rare-event dynamics in complex biomolecular systems. By learning the committor function directly from unbiased MD trajectories, it autonomously identifies the key degrees of freedom governing transitions, and reconstructs transition-state and pathway ensembles without predefined CVs or prior mechanistic knowledge. This strategy enables intricate molecular processes to be captured, such as coupled folding and binding and large-scale conformational changes, where the slowest decorrelating variables are unknown and intrinsically multidimensional. 

Our approach integrates thermodynamics and kinetics within one single framework.
Committor-guided sampling provides access to the TSE and quantitative committor values, thereby linking structural information with kinetic observables.
Reweighting and projecting the sampling onto post-defined CVs yields FELs that offer intuitive visualizations of the underlying processes (see SI). Put together, these features establish Gen-COMPAS as a unified tool for investigating rare events from both energetic and kinetic perspectives.

Unlike traditional enhanced-sampling methods, which accelerate transitions by means of artificial biases, Gen-COMPAS achieves orders-of-magnitude speedups while preserving correct thermodynamics and kinetics. Its efficiency is rooted in an iterative loop of generative modeling and short unbiased simulations. Unlike dataset-based generative approaches, which are limited by the statistical patterns of the pre-collected training data~\cite{jing2024generative, lewis2025scalable, costa2025accelerating}, Gen-COMPAS operates directly on the physical Hamiltonian of the system at hand, requiring no pre-generated datasets and accommodating any molecule or environment, given appropriate force-field parameters. Accuracy can be systematically improved by incorporating higher-fidelity methods, such as ab initio MD or machine-learning force fields. Because sampling is originated from unbiased trajectories, the framework naturally adapts to changes in temperature, pressure, solvent nature, and other environmental factors, ensuring broad applicability and physical grounding.

While Gen-COMPAS provides a remarkably accelerated path to mechanistic discovery, it is best viewed as a computational ecosystem for exploration. The generated FELs represent a powerful tool for identifying key states and barriers, but for applications demanding the highest quantitative precision, they should serve as a well-informed starting point. The true power of the framework lies in its ability to rapidly identify the critical regions of a vast, uncharted conformational space. The TSE can then seed more traditional, computationally intensive techniques, like enhanced sampling or transition path sampling, thereby drastically accelerating the entire pipeline from working hypotheses to rigorously converged results.

In summary, Gen-COMPAS marks a decisive advance toward bridging the longstanding gap between atomistic simulations and mechanistic understanding of complex biomolecular processes. By navigating high-dimensional conformational landscapes with efficiency and precision, and by pinpointing transition states and pathways, it provides an exceptionally rapid route to mechanistic insights. The framework is poised for extension to increasingly complex biological and chemical systems, where refined generative models and committor-guided sampling could capture transitions among multiple metastable states far beyond the simple two-state Markov-jump paradigm. As generative modeling and MD continue to converge, a provocative, albeit legitimate question arises---are we witnessing the beginning of the end for exhaustive $\mu$s-to-ms brute-force simulations and traditional enhanced-sampling strategies? Gen-COMPAS does not entirely replace these approaches, but rather demonstrates that the essence of rare-event dynamics can be revealed without extreme computational cost, potentially reshaping the future trajectory of molecular simulation.





\section*{Acknowledgments}
C.C. acknowledges the European Research Council (project 101097272 ``MilliInMicro''), the Universit\'e de Lorraine through its Lorraine Universit\'e d’Excellence initiative, the R\'egion Grand-Est (project ``Respire''), and the M\'etropole du Grand Nancy (project ``ARC''). M.P.P. acknowledges the funding from the Agence Nationale de la Recherche (ANR, Pestipenta ANR-21-CE11-0016).

\section*{Author contributions}
Data curation and generation: CT. Methodology + conceptualization: CT, MPP, CC. Software: MPP, CT, CGC. Results generation, analysis, validation and visualization: CT, MPP, CGC. Writing: CT, MPP, CGC, AM, FD, CC. Project lead + administration: CC.

\section*{Competing interests}
The authors declare no competing interests.


\clearpage


\clearpage

\begin{thebibliography}{10}
\expandafter\ifx\csname url\endcsname\relax
  \def\url#1{\burl{#1}}\fi
\expandafter\ifx\csname urlprefix\endcsname\relax\def\urlprefix{URL }\fi
\providecommand{\bibinfo}[2]{#2}
\providecommand{\eprint}[2][]{\url{#2}}
\providecommand{\doi}[1]{\url{https://doi.org/#1}}
\bibcommenthead

\bibitem{yang2009}
\bibinfo{author}{Yang, S.}, \bibinfo{author}{Banavali, N.~K.} \& \bibinfo{author}{Roux, B.}
\newblock \bibinfo{title}{Mapping the conformational transition in src activation by cumulating the information from multiple molecular dynamics trajectories}.
\newblock \emph{\bibinfo{journal}{Proc. Natl. Acad. Sci. U.S.A.}} \textbf{\bibinfo{volume}{106}}, \bibinfo{pages}{3776--3781} (\bibinfo{year}{2009}).

\bibitem{Lindorff-Larsen2011fold}
\bibinfo{author}{Lindorff-Larsen, K.}, \bibinfo{author}{Piana, S.}, \bibinfo{author}{Dror, R.~O.} \& \bibinfo{author}{Shaw, D.~E.}
\newblock \bibinfo{title}{{How Fast-Folding Proteins Fold}}.
\newblock \emph{\bibinfo{journal}{Science}} \textbf{\bibinfo{volume}{334}}, \bibinfo{pages}{517--520} (\bibinfo{year}{2011}).

\bibitem{ajaz_2011}
\bibinfo{author}{Ajaz, A.} \emph{et~al.}
\newblock \bibinfo{title}{Concerted vs stepwise mechanisms in dehydro-diels–alder reactions}.
\newblock \emph{\bibinfo{journal}{J. Org. Chem.}} \textbf{\bibinfo{volume}{76}}, \bibinfo{pages}{9320--9328} (\bibinfo{year}{2011}).

\bibitem{moradi2013}
\bibinfo{author}{Moradi, M.} \& \bibinfo{author}{Tajkhorshid, E.}
\newblock \bibinfo{title}{Mechanistic picture for conformational transition of a membrane transporter at atomic resolution}.
\newblock \emph{\bibinfo{journal}{Proc. Natl. Acad. Sci. U.S.A.}} \textbf{\bibinfo{volume}{110}}, \bibinfo{pages}{18916--18921} (\bibinfo{year}{2013}).

\bibitem{BK01}
\bibinfo{author}{Biroli, G.} \& \bibinfo{author}{Kurchan, J.}
\newblock \bibinfo{title}{Metastable states in glassy systems}.
\newblock \emph{\bibinfo{journal}{Phys. Rev. E}} \textbf{\bibinfo{volume}{64}}, \bibinfo{pages}{016101} (\bibinfo{year}{2001}).

\bibitem{sipka_differentiable_2023_inproceedings}
\bibinfo{author}{Sipka, M.}, \bibinfo{author}{Dietschreit, J. C.~B.}, \bibinfo{author}{Grajciar, L.} \& \bibinfo{author}{Gomez-Bombarelli, R.}
\newblock \bibinfo{title}{Differentiable simulations for enhanced sampling of rare events}.
\newblock \emph{\bibinfo{booktitle}{Proc. International Conference on Learning Representations (ICLR)}}, \bibinfo{pages}{31990--32007} (\bibinfo{publisher}{PMLR}, \bibinfo{year}{2023}).

\bibitem{jung2023machine}
\bibinfo{author}{Jung, H.} \emph{et~al.}
\newblock \bibinfo{title}{Machine-guided path sampling to discover mechanisms of molecular self-organization}.
\newblock \emph{\bibinfo{journal}{Nat. Comput. Sci.}} \textbf{\bibinfo{volume}{3}}, \bibinfo{pages}{334--345} (\bibinfo{year}{2023}).

\bibitem{kang2024Parrinello}
\bibinfo{author}{Kang, P.}, \bibinfo{author}{Trizio, E.} \& \bibinfo{author}{Parrinello, M.}
\newblock \bibinfo{title}{Computing the committor with the committor to study the transition state ensemble}.
\newblock \emph{\bibinfo{journal}{Nat. Comput. Sci.}} \textbf{\bibinfo{volume}{4}}, \bibinfo{pages}{451–460} (\bibinfo{year}{2024}).

\bibitem{Dror2012}
\bibinfo{author}{Dror, R.~O.}, \bibinfo{author}{Dirks, R.~M.}, \bibinfo{author}{Grossman, J.~P.}, \bibinfo{author}{Xu, H.} \& \bibinfo{author}{Shaw, D.~E.}
\newblock \bibinfo{title}{Biomolecular simulation: A computational microscope for molecular biology}.
\newblock \emph{\bibinfo{journal}{Annu. Rev. Biophys.}} \textbf{\bibinfo{volume}{41}}, \bibinfo{pages}{429--452} (\bibinfo{year}{2012}).

\bibitem{Chipot2023}
\bibinfo{author}{Chipot, C.}
\newblock \bibinfo{title}{Free energy methods for the description of molecular processes}.
\newblock \emph{\bibinfo{journal}{Ann. Rev. Biophys.}} \textbf{\bibinfo{volume}{52}}, \bibinfo{pages}{113--138} (\bibinfo{year}{2023}).

\bibitem{rogal_reaction_2021}
\bibinfo{author}{Rogal, J.}
\newblock \bibinfo{title}{Reaction coordinates in complex systems-a perspective}.
\newblock \emph{\bibinfo{journal}{Eur. Phys. J. B}} \textbf{\bibinfo{volume}{94}}, \bibinfo{pages}{223} (\bibinfo{year}{2021}).

\bibitem{branduardi_b_2007}
\bibinfo{author}{Branduardi, D.}, \bibinfo{author}{Gervasio, F.~L.} \& \bibinfo{author}{Parrinello, M.}
\newblock \bibinfo{title}{From {A} to {B} in free energy space}.
\newblock \emph{\bibinfo{journal}{J. Chem. Phys.}} \textbf{\bibinfo{volume}{126}}, \bibinfo{pages}{054103} (\bibinfo{year}{2007}).

\bibitem{chen_enhancing_2022}
\bibinfo{author}{Chen, H.} \& \bibinfo{author}{Chipot, C.}
\newblock \bibinfo{title}{Enhancing sampling with free-energy calculations}.
\newblock \emph{\bibinfo{journal}{Curr. Opin. Struct. Biol.}} \textbf{\bibinfo{volume}{77}}, \bibinfo{pages}{102497} (\bibinfo{year}{2022}).

\bibitem{shaw2008anton}
\bibinfo{author}{Shaw, D.~E.} \emph{et~al.}
\newblock \bibinfo{title}{Anton, a special-purpose machine for molecular dynamics simulation}.
\newblock \emph{\bibinfo{journal}{Communications of the ACM}} \textbf{\bibinfo{volume}{51}}, \bibinfo{pages}{91--97} (\bibinfo{year}{2008}).

\bibitem{Shaw2021}
\bibinfo{author}{Shaw, D.~E.} \emph{et~al.}
\newblock \bibinfo{title}{Anton 3: Twenty microseconds of molecular dynamics simulation before lunch}.
\newblock \emph{\bibinfo{booktitle}{{SC} '21: The International Conference for High Performance Computing, Networking, Storage and Analysis, St. Louis, Missouri, USA, November 14 - 19, 2021}}, \bibinfo{pages}{1:1--1:11} (\bibinfo{year}{2021}).

\bibitem{WN18}
\bibinfo{author}{Wehmeyer, C.} \& \bibinfo{author}{Noé, F.}
\newblock \bibinfo{title}{Time-lagged autoencoders: Deep learning of slow collective variables for molecular kinetics}.
\newblock \emph{\bibinfo{journal}{J. Chem. Phys.}} \textbf{\bibinfo{volume}{148}}, \bibinfo{pages}{241703} (\bibinfo{year}{2018}).

\bibitem{BGLS21}
\bibinfo{author}{Belkacemi, Z.}, \bibinfo{author}{Gkeka, P.}, \bibinfo{author}{Lelièvre, T.} \& \bibinfo{author}{Stoltz, G.}
\newblock \bibinfo{title}{Chasing collective variables using autoencoders and biased trajectories}.
\newblock \emph{\bibinfo{journal}{J. Chem. Theory Comput.}} \textbf{\bibinfo{volume}{18}}, \bibinfo{pages}{59--78} (\bibinfo{year}{2022}).

\bibitem{FAB21}
\bibinfo{author}{Frassek, M.}, \bibinfo{author}{Arjun, A.} \& \bibinfo{author}{Bolhuis, P.~G.}
\newblock \bibinfo{title}{An extended autoencoder model for reaction coordinate discovery in rare event molecular dynamics datasets}.
\newblock \emph{\bibinfo{journal}{J. Chem. Phys.}} \textbf{\bibinfo{volume}{155}}, \bibinfo{pages}{064103} (\bibinfo{year}{2021}).

\bibitem{Pengmei2025}
\bibinfo{author}{Pengmei, Z.}, \bibinfo{author}{Lorpaiboon, C.}, \bibinfo{author}{Guo, S.~C.}, \bibinfo{author}{Weare, J.} \& \bibinfo{author}{Dinner, A.~R.}
\newblock \bibinfo{title}{Using pretrained graph neural networks with token mixers as geometric featurizers for conformational dynamics}.
\newblock \emph{\bibinfo{journal}{J. Chem. Phys.}} \textbf{\bibinfo{volume}{162}}, \bibinfo{pages}{044107} (\bibinfo{year}{2025}).

\bibitem{ZWT25}
\bibinfo{author}{Zou, Z.}, \bibinfo{author}{Wang, D.} \& \bibinfo{author}{Tiwary, P.}
\newblock \bibinfo{title}{A graph neural network-state predictive information bottleneck (gnn-spib) approach for learning molecular thermodynamics and kinetics}.
\newblock \emph{\bibinfo{journal}{Digit. Discov.}} \textbf{\bibinfo{volume}{4}}, \bibinfo{pages}{211--221} (\bibinfo{year}{2025}).

\bibitem{Miao2021}
\bibinfo{author}{Miao, M.} \emph{et~al.}
\newblock \bibinfo{title}{Avoiding non-equilibrium effects in adaptive biasing force calculations}.
\newblock \emph{\bibinfo{journal}{Mol. Simul.}} \textbf{\bibinfo{volume}{47}}, \bibinfo{pages}{390--394} (\bibinfo{year}{2021}).

\bibitem{e_finite_2005}
\bibinfo{author}{E, W.}, \bibinfo{author}{Ren, W.} \& \bibinfo{author}{Vanden-Eijnden, E.}
\newblock \bibinfo{title}{Finite temperature string method for the study of rare events}.
\newblock \emph{\bibinfo{journal}{J. Phys. Chem. B}} \textbf{\bibinfo{volume}{109}}, \bibinfo{pages}{6688--6693} (\bibinfo{year}{2005}).

\bibitem{maragliano2006string}
\bibinfo{author}{Maragliano, L.}, \bibinfo{author}{Fischer, A.}, \bibinfo{author}{Vanden-Eijnden, E.} \& \bibinfo{author}{Ciccotti, G.}
\newblock \bibinfo{title}{{String method in collective variables: Minimum free energy paths and isocommittor surfaces}}.
\newblock \emph{\bibinfo{journal}{J. Chem. Phys.}} \textbf{\bibinfo{volume}{125}}, \bibinfo{pages}{024106} (\bibinfo{year}{2006}).

\bibitem{fu2020finding}
\bibinfo{author}{Fu, H.} \emph{et~al.}
\newblock \bibinfo{title}{Finding an optimal pathway on a multidimensional free-energy landscape}.
\newblock \emph{\bibinfo{journal}{J. Chem. Inf. Model.}} \textbf{\bibinfo{volume}{60}}, \bibinfo{pages}{5366--5374} (\bibinfo{year}{2020}).

\bibitem{pan_finding_2008}
\bibinfo{author}{Pan, A.~C.}, \bibinfo{author}{Sezer, D.} \& \bibinfo{author}{Roux, B.}
\newblock \bibinfo{title}{Finding transition pathways using the string method with swarms of trajectories}.
\newblock \emph{\bibinfo{journal}{J. Phys. Chem. B}} \textbf{\bibinfo{volume}{112}}, \bibinfo{pages}{3432--3440} (\bibinfo{year}{2008}).

\bibitem{VT05}
\bibinfo{author}{Vanden-Eijnden, E.} \& \bibinfo{author}{Tal, F.~A.}
\newblock \bibinfo{title}{Transition state theory: Variational formulation, dynamical corrections, and error estimates}.
\newblock \emph{\bibinfo{journal}{J. Chem. Phys.}} \textbf{\bibinfo{volume}{123}}, \bibinfo{pages}{184103} (\bibinfo{year}{2005}).

\bibitem{SPNCB19}
\bibinfo{author}{Swenson, D. W.~H.}, \bibinfo{author}{Prinz, J.-H.}, \bibinfo{author}{No{\'e}, F.}, \bibinfo{author}{Chodera, J.~D.} \& \bibinfo{author}{Bolhuis, P.~G.}
\newblock \bibinfo{title}{{OpenPathSampling}: {A} {Python} framework for path sampling simulations. 1. {Basics}}.
\newblock \emph{\bibinfo{journal}{J. Chem. Theory Comput.}} \textbf{\bibinfo{volume}{15}}, \bibinfo{pages}{813--836} (\bibinfo{year}{2019}).

\bibitem{roux2022transition}
\bibinfo{author}{Roux, B.}
\newblock \bibinfo{title}{{Transition rate theory, spectral analysis, and reactive paths}}.
\newblock \emph{\bibinfo{journal}{J. Chem. Phys}} \textbf{\bibinfo{volume}{156}}, \bibinfo{pages}{134111} (\bibinfo{year}{2022}).

\bibitem{strahan2023predicting}
\bibinfo{author}{Strahan, J.}, \bibinfo{author}{Finkel, J.}, \bibinfo{author}{Dinner, A.~R.} \& \bibinfo{author}{Weare, J.}
\newblock \bibinfo{title}{Predicting rare events using neural networks and short-trajectory data}.
\newblock \emph{\bibinfo{journal}{J. Comput. Phys.}} \textbf{\bibinfo{volume}{488}}, \bibinfo{pages}{112152} (\bibinfo{year}{2023}).

\bibitem{he_committor-consistent_2022}
\bibinfo{author}{He, Z.}, \bibinfo{author}{Chipot, C.} \& \bibinfo{author}{Roux, B.}
\newblock \bibinfo{title}{Committor-consistent variational string method}.
\newblock \emph{\bibinfo{journal}{J. Phys. Chem. Lett.}} \textbf{\bibinfo{volume}{13}}, \bibinfo{pages}{9263--9271} (\bibinfo{year}{2022}).

\bibitem{chen_discovering_2023}
\bibinfo{author}{Chen, H.}, \bibinfo{author}{Roux, B.} \& \bibinfo{author}{Chipot, C.}
\newblock \bibinfo{title}{Discovering reaction pathways, slow variables, and committor probabilities with machine learning}.
\newblock \emph{\bibinfo{journal}{J. Chem. Theory Comput.}} \textbf{\bibinfo{volume}{19}}, \bibinfo{pages}{4414--4426} (\bibinfo{year}{2023}).

\bibitem{MCCCRC25}
\bibinfo{author}{Meg\'ias, A.} \emph{et~al.}
\newblock \bibinfo{title}{Iterative variational learning of committor-consistent transition pathways using artificial neural networks}.
\newblock \emph{\bibinfo{journal}{Nat. Comput. Sci.}} \textbf{\bibinfo{volume}{5}}, \bibinfo{pages}{592–602} (\bibinfo{year}{2025}).

\bibitem{giuseppe_grad_2025}
\bibinfo{author}{Chen, C.~G.} \emph{et~al.}
\newblock \bibinfo{title}{Following the committor flow: A data-driven discovery of transition pathways}.
\newblock \emph{\bibinfo{journal}{arXiv preprint arXiv:2507.21961}}  (\bibinfo{year}{2025}).

\bibitem{sergio_qGNN_2025}
\bibinfo{author}{Arredondo, S.~C.} \emph{et~al.}
\newblock \bibinfo{title}{From atoms to dynamics: Learning the committor without collective variables}.
\newblock \emph{\bibinfo{journal}{arXiv preprint arXiv:2507.17700}}  (\bibinfo{year}{2025}).

\bibitem{noe2019boltzmann}
\bibinfo{author}{No{\'e}, F.}, \bibinfo{author}{Olsson, S.}, \bibinfo{author}{K{\"o}hler, J.} \& \bibinfo{author}{Wu, H.}
\newblock \bibinfo{title}{Boltzmann generators: Sampling equilibrium states of many-body systems with deep learning}.
\newblock \emph{\bibinfo{journal}{Science}} \textbf{\bibinfo{volume}{365}}, \bibinfo{pages}{eaaw1147} (\bibinfo{year}{2019}).

\bibitem{jing2024generative}
\bibinfo{author}{Jing, B.}, \bibinfo{author}{St{\"a}rk, H.}, \bibinfo{author}{Jaakkola, T.} \& \bibinfo{author}{Berger, B.}
\newblock \bibinfo{title}{Generative modeling of molecular dynamics trajectories}.
\newblock \emph{\bibinfo{booktitle}{Proc. Neural Information Processing Systems (NeurIPS)}}, Vol.~\bibinfo{volume}{37}, \bibinfo{pages}{40534--40564} (\bibinfo{year}{2024}).

\bibitem{lewis2025scalable}
\bibinfo{author}{Lewis, S.} \emph{et~al.}
\newblock \bibinfo{title}{Scalable emulation of protein equilibrium ensembles with generative deep learning}.
\newblock \emph{\bibinfo{journal}{Science}} \textbf{\bibinfo{volume}{389}}, \bibinfo{pages}{eadv9817} (\bibinfo{year}{2025}).

\bibitem{ho2020denoising}
\bibinfo{author}{Ho, J.}, \bibinfo{author}{Jain, A.} \& \bibinfo{author}{Abbeel, P.}
\newblock \bibinfo{title}{Denoising diffusion probabilistic models}.
\newblock \emph{\bibinfo{booktitle}{Proc. Neural Information Processing Systems (NeurIPS)}}, Vol.~\bibinfo{volume}{33}, \bibinfo{pages}{6840--6851} (\bibinfo{year}{2020}).

\bibitem{song2020denoising}
\bibinfo{author}{Song, J.}, \bibinfo{author}{Meng, C.} \& \bibinfo{author}{Ermon, S.}
\newblock \bibinfo{title}{Denoising diffusion implicit models}.
\newblock \emph{\bibinfo{booktitle}{Proc. International Conference on Learning Representations (ICLR)}} (\bibinfo{year}{2021}).

\bibitem{Schlitter1994}
\bibinfo{author}{Schlitter, J.}, \bibinfo{author}{Engels, M.} \& \bibinfo{author}{Kr\"uger, P.}
\newblock \bibinfo{title}{Targeted molecular dynamics: A new approach for searching pathways of conformational transitions}.
\newblock \emph{\bibinfo{journal}{J. Mol. Graph.}} \textbf{\bibinfo{volume}{12}}, \bibinfo{pages}{84--89} (\bibinfo{year}{1994}).

\bibitem{costa2025accelerating}
\bibinfo{author}{Costa, A. d.~S.}, \bibinfo{author}{Ponnapati, M.}, \bibinfo{author}{Rubin, D.}, \bibinfo{author}{Smidt, T.} \& \bibinfo{author}{Jacobson, J.}
\newblock \bibinfo{title}{Accelerating protein molecular dynamics simulation with deepjump}.
\newblock \emph{\bibinfo{journal}{arXiv preprint arXiv:2509.13294}}  (\bibinfo{year}{2025}).

\bibitem{chen_chasing_2023}
\bibinfo{author}{Chen, H.} \& \bibinfo{author}{Chipot, C.}
\newblock \bibinfo{title}{Chasing collective variables using temporal data-driven strategies}.
\newblock \emph{\bibinfo{journal}{QRB Discov.}} \textbf{\bibinfo{volume}{4}}, \bibinfo{pages}{e2} (\bibinfo{year}{2023}).

\bibitem{chen_companion_2022}
\bibinfo{author}{Chen, H.} \emph{et~al.}
\newblock \bibinfo{title}{A companion guide to the string method with swarms of trajectories: Characterization, performance, and pitfalls}.
\newblock \emph{\bibinfo{journal}{J. Chem. Theory Comput.}} \textbf{\bibinfo{volume}{18}}, \bibinfo{pages}{1406--1422} (\bibinfo{year}{2022}).

\bibitem{Zhou2003_FEL_TrpCage}
\bibinfo{author}{Zhou, R.}
\newblock \bibinfo{title}{Trp-cage: Folding free energy landscape in explicit water}.
\newblock \emph{\bibinfo{journal}{Proc. Natl. Acad. Sci. U. S. A.}} \textbf{\bibinfo{volume}{100}}, \bibinfo{pages}{13280–13285} (\bibinfo{year}{2003}).

\bibitem{Marinelli2009_KineticModel_TrpCage}
\bibinfo{author}{Marinelli, F.}, \bibinfo{author}{Pietrucci, F.}, \bibinfo{author}{Laio, A.} \& \bibinfo{author}{Piana, S.}
\newblock \bibinfo{title}{A kinetic model of trp-cage folding from multiple biased molecular dynamics simulations}.
\newblock \emph{\bibinfo{journal}{PLoS Comput. Biol.}} \textbf{\bibinfo{volume}{5}}, \bibinfo{pages}{e1000452} (\bibinfo{year}{2009}).

\bibitem{Sidky2019_SRV_MSM_TrpCage}
\bibinfo{author}{Sidky, H.}, \bibinfo{author}{Chen, W.} \& \bibinfo{author}{Ferguson, A.~L.}
\newblock \bibinfo{title}{High-resolution markov state models for the dynamics of trp-cage miniprotein constructed over slow folding modes identified by state-free reversible vampnets}.
\newblock \emph{\bibinfo{journal}{J. Phys. Chem. B}} \textbf{\bibinfo{volume}{123}}, \bibinfo{pages}{7999--8009} (\bibinfo{year}{2019}).

\bibitem{Juraszek2006}
\bibinfo{author}{Juraszek, J.} \& \bibinfo{author}{Bolhuis, P.~G.}
\newblock \bibinfo{title}{Sampling the multiple folding mechanisms of trp-cage in explicit solvent}.
\newblock \emph{\bibinfo{journal}{Proc. Natl. Acad. Sci. U.S.A.}} \textbf{\bibinfo{volume}{103}}, \bibinfo{pages}{15859--15864} (\bibinfo{year}{2006}).

\bibitem{ravindranathan_ribose_2005}
\bibinfo{author}{Ravindranathan, K.~P.}, \bibinfo{author}{Gallicchio, E.} \& \bibinfo{author}{Levy, R.~M.}
\newblock \bibinfo{title}{Conformational equilibria and free energy profiles for the allosteric transition of the ribose-binding protein}.
\newblock \emph{\bibinfo{journal}{J. Mol. Biol.}} \textbf{\bibinfo{volume}{353}}, \bibinfo{pages}{196--210} (\bibinfo{year}{2005}).

\bibitem{Li2013_RBP_ENM}
\bibinfo{author}{Li, H.~Y.}, \bibinfo{author}{Cao, Z.~X.}, \bibinfo{author}{Zhao, L.~L.} \& \bibinfo{author}{Wang, J.~H.}
\newblock \bibinfo{title}{Analysis of conformational motions and residue fluctuations for escherichia coli ribose-binding protein revealed with elastic network models}.
\newblock \emph{\bibinfo{journal}{Int. J. Mol. Sci.}} \textbf{\bibinfo{volume}{14}}, \bibinfo{pages}{10552--10569} (\bibinfo{year}{2013}).

\bibitem{Ren2021_RBP_G134R_Coupling}
\bibinfo{author}{Ren, W.} \emph{et~al.}
\newblock \bibinfo{title}{Unraveling the coupling between conformational changes and ligand binding in rbp\_g134r via enhanced sampling}.
\newblock \emph{\bibinfo{journal}{J. Phys. Chem. B}} \textbf{\bibinfo{volume}{125}}, \bibinfo{pages}{2898--2909} (\bibinfo{year}{2021}).

\bibitem{Li2021_RBP_LigandInduced_MD}
\bibinfo{author}{Li, H.} \emph{et~al.}
\newblock \bibinfo{title}{Ligand-induced structural changes analysis of rbp as studied by md simulations}.
\newblock \emph{\bibinfo{journal}{Technol. Health Care}} \textbf{\bibinfo{volume}{29}}, \bibinfo{pages}{103--114} (\bibinfo{year}{2021}).

\bibitem{klingenberg_aac_2008}
\bibinfo{author}{Klingenberg, M.}
\newblock \bibinfo{title}{The adp and atp transport in mitochondria and its carrier}.
\newblock \emph{\bibinfo{journal}{Biochim. Biophys. Acta, Biomembr.}} \textbf{\bibinfo{volume}{1778}}, \bibinfo{pages}{1978--2021} (\bibinfo{year}{2008}).

\bibitem{mavridou_membrane_2024}
\bibinfo{author}{Mavridou, V.}, \bibinfo{author}{King, M.~S.}, \bibinfo{author}{Bazzone, A.}, \bibinfo{author}{Springett, R.} \& \bibinfo{author}{Kunji, E. R.~S.}
\newblock \bibinfo{title}{Membrane potential stimulates adp import and atp export by the mitochondrial {ADP}/{ATP} carrier due to its positively charged binding site}.
\newblock \emph{\bibinfo{journal}{Sci. Adv.}} \textbf{\bibinfo{volume}{10}}, \bibinfo{pages}{eadp7725} (\bibinfo{year}{2024}).

\bibitem{kunji_aac_2020}
\bibinfo{author}{Kunji, E.~R.} \& \bibinfo{author}{Ruprecht, J.~J.}
\newblock \bibinfo{title}{The mitochondrial {ADP}/{ATP} carrier exists and functions as a monomer}.
\newblock \emph{\bibinfo{journal}{Biochem. Soc. Trans.}} \textbf{\bibinfo{volume}{48}}, \bibinfo{pages}{1419--1432} (\bibinfo{year}{2020}).

\bibitem{senoo_functional_2024}
\bibinfo{author}{Senoo, N.} \emph{et~al.}
\newblock \bibinfo{title}{Functional diversity among cardiolipin binding sites on the mitochondrial {ADP}/{ATP} carrier}.
\newblock \emph{\bibinfo{journal}{EMBO J.}} \textbf{\bibinfo{volume}{43}}, \bibinfo{pages}{2979--3008} (\bibinfo{year}{2024}).

\bibitem{tamura_atomistic_2017}
\bibinfo{author}{Tamura, K.} \& \bibinfo{author}{Hayashi, S.}
\newblock \bibinfo{title}{Atomistic modeling of alternating access of a mitochondrial {ADP}/{ATP} membrane transporter with molecular simulations}.
\newblock \emph{\bibinfo{journal}{PLOS ONE}} \textbf{\bibinfo{volume}{12}}, \bibinfo{pages}{1--21} (\bibinfo{year}{2017}).

\bibitem{skulj_molecular_2020}
\bibinfo{author}{Škulj, S.}, \bibinfo{author}{Brkljača, Z.} \& \bibinfo{author}{Vazdar, M.}
\newblock \bibinfo{title}{Molecular dynamics simulations of the elusive matrix-open state of mitochondrial {ADP}/{ATP} carrier}.
\newblock \emph{\bibinfo{journal}{Isr. J. Chem.}} \textbf{\bibinfo{volume}{60}}, \bibinfo{pages}{735--743} (\bibinfo{year}{2020}).

\bibitem{quadrotta_modeling_2025}
\bibinfo{author}{Quadrotta, V.} \& \bibinfo{author}{Polticelli, F.}
\newblock \bibinfo{title}{Modeling the different conformations of the human mitochondrial {ADP}/{ATP} carrier using alphafold and molecular dynamics simulations of the protein-ligand complexes}.
\newblock \emph{\bibinfo{journal}{Comput. Struct. Biotechnol. J.}} \textbf{\bibinfo{volume}{27}}, \bibinfo{pages}{1265--1277} (\bibinfo{year}{2025}).

\bibitem{springett_modelling_2017}
\bibinfo{author}{Springett, R.}, \bibinfo{author}{King, M.~S.}, \bibinfo{author}{Crichton, P.~G.} \& \bibinfo{author}{Kunji, E.~R.}
\newblock \bibinfo{title}{Modelling the free energy profile of the mitochondrial {ADP}/{ATP} carrier}.
\newblock \emph{\bibinfo{journal}{Biochim. Biophys. Acta, Bioenerg.}} \textbf{\bibinfo{volume}{1858}}, \bibinfo{pages}{906--914} (\bibinfo{year}{2017}).

\bibitem{pietropaolo_switching_2016}
\bibinfo{author}{Pietropaolo, A.}, \bibinfo{author}{Pierri, C.~L.}, \bibinfo{author}{Palmieri, F.} \& \bibinfo{author}{Klingenberg, M.}
\newblock \bibinfo{title}{The switching mechanism of the mitochondrial {ADP}/{ATP} carrier explored by free-energy landscapes}.
\newblock \emph{\bibinfo{journal}{Biochim. Biophys. Acta, Bioenerg.}} \textbf{\bibinfo{volume}{1857}}, \bibinfo{pages}{772--781} (\bibinfo{year}{2016}).

\bibitem{yao_computational_2022}
\bibinfo{author}{Yao, S.} \emph{et~al.}
\newblock \bibinfo{title}{Mechanistic insights into multiple-step transport of mitochondrial {ADP}/{ATP} carrier}.
\newblock \emph{\bibinfo{journal}{Comput. Struct. Biotechnol. J.}} \textbf{\bibinfo{volume}{20}}, \bibinfo{pages}{1829--1840} (\bibinfo{year}{2022}).

\bibitem{yi_molecular_2019}
\bibinfo{author}{Yi, Q.} \emph{et~al.}
\newblock \bibinfo{title}{Molecular dynamics simulations on apo {ADP}/{ATP} carrier shed new lights on the featured motif of the mitochondrial carriers}.
\newblock \emph{\bibinfo{journal}{Mitochondrion}} \textbf{\bibinfo{volume}{47}}, \bibinfo{pages}{94--102} (\bibinfo{year}{2019}).

\bibitem{kurauskas_dynamics_2018}
\bibinfo{author}{Kurauskas, V.} \emph{et~al.}
\newblock \bibinfo{title}{Dynamics and interactions of {AAC3} in {DPC} are not functionally relevant}.
\newblock \emph{\bibinfo{journal}{Nat. Struct. Mol. Biol.}} \textbf{\bibinfo{volume}{25}}, \bibinfo{pages}{745--747} (\bibinfo{year}{2018}).

\end{thebibliography}
\end{document}